\documentclass[preprint]{revtex4-1}

\usepackage{graphicx,color}
\usepackage{amssymb}

\begin{document}


\title{
Waveguide-mode interference lithography technique for high contrast subwavelength structures in the visible region
}


\author{
Kanta Kusaka, Hiroyuki Kurosawa, Seigo Ohno$^*$,Yozaburo Sakaki,\\ Kazuyuki Nakayama, Yuto Moritake, and Teruya Ishihara\\}
\affiliation{Department of Physics, Graduate School of Science, Tohoku University, Sendai, 
980-8578, Japan\\
$^*$seigo@m.tohoku.ac.jp
}


\date{\today}

\begin{abstract}
We explore possibilities of waveguide-mode interference lithography (WMIL) technique for high contrast subwavelength structures in the visible region. Selecting an appropriate waveguide-mode, we demonstrate high contrast resist mask patterns for the first time. TM1 mode in the waveguide is shown to be useful for providing a three-dimensional structure whose cross section is checkerboard pattern. Applying our WMIL technique, we demonstrate 1D, 2D and 3D subwavelength resist patterns that are widely used for the fabrication of metamteterials in the visible region. In addition to the resist patterns, we demonstrate a resonance at 1.9 eV for a split tube structure experimentally.
\end{abstract}


\maketitle

\section{Introduction}
Metamaterials whose characteristic structure scale is in subwavelength regime have attracted much attention due to its potential to design the optical properties and their exotic behavior that cannot be available in nature. In the early days of metameterial studies the fundamental principles of the exotic behavior were investigated in the microwave region [1,2]. Application of the nanotechnology for semiconductor processing makes the feature size of metamaterials as small as a few hundred nanometers, with which they can operate in the visible region [3-5]. Electron beam (EB) lithography or focused ion beam (FIB) techniques have been applied to nano-size fabrication of metamaterial components.  The typical area size of 100 $\mu$m x 100 $\mu$m fabricated by these techniques is suitable for the trial production, but is not enough for practical applications or angle resolved measurements. 

Photolithography techniques, on the other hand, have an advantage for high-speed and large areas processing. Especially the interference lithography is a simple and cost-effective way to pattern a periodic photo-resist without a mask. As for the resolution of the photolithography, it is usually subject to the diffraction limit of light. Therefore a shorter wavelength light source and corresponding photo-resist materials are required in order to achieve fine resolution enough for the subwavelength structure. Luo and Ishihara improved an evanescent interference lithography proposed by Blaikie and McNab [6] and demonstrated subwavelength structure prepared by surface plasmon resonant interference nanolithography technique (SPRINT), which can overcome the diffraction limit due to large wavevector inherent to the plasmonic nature [7]. Although they used metallic periodic structure in order to excite surface plasmon, maskless interference lithography using attenuated total reflection geometry is also demonstrated to form fine structures in a photo-resist [8]. However, since the near field light in the photo-resist decays exponentially along the thickness direction, it is not suitable for fabricating a structure with high-aspect ratio, which is frequently required for dry etching processes. 

Recently Wang et al. succeeded to fabricate sub-wavelength structures by utilizing waveguide modes in an interference lithography technique [9]. They interfered two laser beams in a photoresist waveguide coupled with a prism through a thin metal layer. Although they pointed out that the method has an advantage for thicker resist and suggested the usage of higher modes, their demonstrations were limited in relatively simpler cases. Here we point out that the most significant difference between the conventional and the waveguide interference lithographies resides in the crisp intensity distribution of the latter, which comes from the standing wave nature of the interference. In the presenting paper, we explore possibilities of waveguide-mode interference lithography (WMIL) technique, aiming for simple fabrication of metamaterials in the visible region.

In the next section we show our experimental setup for WMIL. We introduce several configurations of WMIL in the following sections: In section 3, we introduce additional structure in the dielectric layer. In section 4, we introduce additional periodicity in the orthogonal direction to form two-dimensional structures on substrates. In section 5, we introduce further structure along the thickness direction. In section 6, we replace a single metal layer with metal / dielectric multilayers. In section 7, we show that a metamaterial having resonance in the visible region can be fabricated by applying WMIL. Finally, we conclude our work in the last section.

\section{Optical setup for WMIL and sample preparation}
Our experimental setup for WMIL is schematically shown in Fig. 1. This configuration is basically similar to that in Ref. [9].  In our setup we used a single mode fiber in order to obtain high beam quality and to introduce a beam to an optical breadboard isolated from the vibration due to the cooling fan of the laser. The optics before the single mode fiber are used to control the coupling efficiency of the fiber. The beam output from the fiber is split into two arms by a polarization beam splitter (PBS) (Thorlabs, PBS251), which are used for an interference exposure. Half-wave plates are inserted in each arm to control the polarization. Besides a quarter-wave plate is inserted to compensate the deviation from a linear polarization of the reflected beam from the PBS. The split beams are incident on a high refractive index prism and couple to a photoresist waveguide, where interference forms spatial modulation of light intensity. For our purpose of subwavelength structures in the visible region by WMIL, we chose a higher refractive index material of SF-11, n = 1.83 at 442 nm [10] for the prism and the sample substrate in order to shorten the period of interference patterns. 
 \begin{figure}
\includegraphics[width=0.6\linewidth]{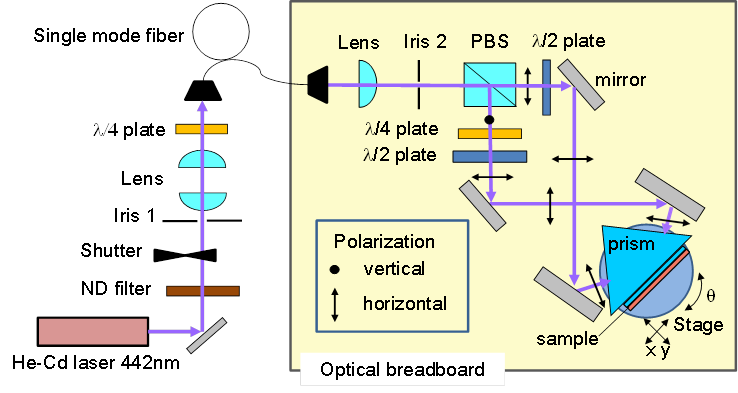}%
\caption{Experimental setup for the waveguide mode interference lithography. PBS denotes Polarization Beam Splitter. The polarization direction is corresponding to using the TM modes.\label{fig1.png}}
 \end{figure}

The sample preparation process is summarized as follows. First, a metallic thin film is prepared on SF-11 (n-SF-11, HOYA) substrate by an ion beam sputtering technique (EIS-230S8, ELIONIX), whose area and thickness are 15 mm x 15 mm and 0.5 mm, respectively. Then, a dielectric waveguide is formed on the substrate with metallic film by spin-coating photo-resist (S1813, SHIPLEY). Thicknesses of these layers will be specified in each section.

In the exposure process, an unexposed sample is attached on the prism with an index-matching liquid (Shimadzu Device Corporation), which is used for an optical contact. Since it is not practical to find the resonance condition from the geometrical parameters, we varied the incidence angle and measured the reflectance from the sample surface. Then we determined the resonance incident angle in the case where the reflected light intensity is minimum. When we find the resonance incident angle, the diameter of the beam is set to be 1 mm by the iris 2 shown in Fig. 1 to limit the sacrificed area by the exposure. Then the interference exposure is carried out using a full diameter of the beams, which was 10 mm in our setup. After the exposure we developed the sample with a solution containing sodium hydroxide according to the standard prescription.

\section{Modifying dielectric waveguide layer}
It is known that multi-dielectric layers made of different materials work as a waveguide [11]. Here we consider the case where the dielectric waveguide in WMIL consists of multi-layers. Such situation in a waveguide is interesting from two points of view. The first is the controllability of the electric field distribution in the waveguide. The distribution of refractive index modulates electric field in the waveguide. The electric field at the high index enhanced higher. Secondly, by inserting a transparent dielectric layer between a metal coupler and photo-resist layer, it is possible to use chemically unstable metals. For example, An Al film can be protected against alkaline resist developer. A Ag film can be also protected from oxidization in the air. 

An example of a bilayer waveguide is shown in Fig.2 (a). In this design, a thin alumina layer is inserted between a silver layer and a polymer layer. The simulated waveguide mode TE1 shown in Fig. 2 (b) is resonant at the incident angle of $\theta$ = 57.0$^\circ$. Because the thickness of alumina layer of 20 nm is thin and that the refractive index of n$_{\rm Al_2O_3}$ = 1.77 is quite close to that of polymer of n$_{\rm S1813}$ = 1.69, the electric field distribution in the bilayer waveguide is almost identical with that of a single layer. However, the thickness of alumina layer is sufficiently thick to prevent the metal layer from oxidizing in the air. Adopting this design we fabricated a sample by the lithography process mentioned in the previous section.Figures 2 (c) and 2 (d) are AFM images for front and birdfs eye view of the fabricated sample, respectively. The cross-section of the grating is shown in Fig. 2 (e). The period and height of the structure were measured as 165 nm and 158 nm, respectively. The aspect ratio of this sample is around 1 and rectangular shape of each structure is confirmed from the figure. The vertical wall in a structure is one of the characteristic features of WMIL.

 \begin{figure}
\includegraphics[width=0.6\linewidth]{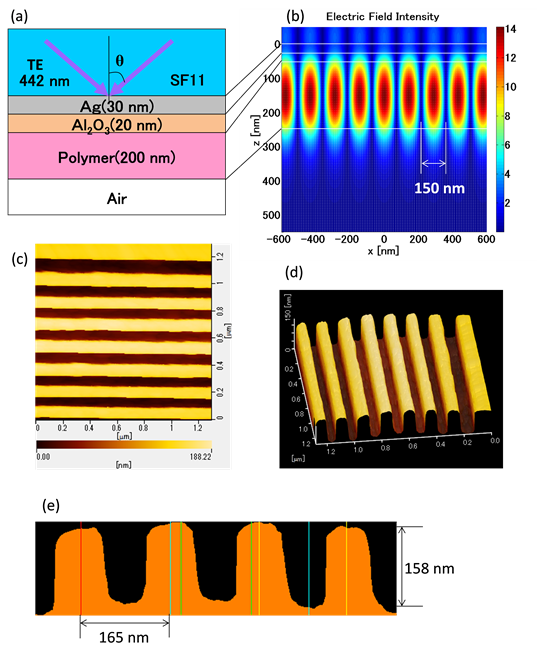}%
\caption{ Example of a bilayered waveguide sample. Configuration of simulation (a) and simulated results of electric field distribution (b). AFM image of top view (c) and bird-eye view and the results of its cross-section analysis (e).\label{fig1.png}}
 \end{figure}

\section{Two-dimensional structures on substrate}
By combining the waveguide mode lithography with a multiple exposure technique we can easily fabricate the two-dimensional structure on a substrate. Here, we investigate two types of structure of dot matrix and fishnet.
By combining the waveguide mode lithography with a multiple exposure technique we can easily fabricate two-dimensional structures on a substrate. Here, we investigate two types of structure of dot matrix and fishnet.

For the fabrication of such structures we designed a waveguide shown in Fig. 3(a) which consists of a Ag layer (20nm) and a photo-resist layer (200 nm) on a substrate of SF11. TE1 mode with a period of 150 nm can be excited when the incident angle ? is 55.3o.
 
To fabricate a two-dimensional structure on the substrate, we exposed a sample twice. After the first exposure, we rotated the sample by 90 degrees on a prism and then exposed again. By controlling the exposing time we can change the developed structure in the exposing process. The schematic configuration of the exposing process is shown in Fig. 3 (b).  In the figure, the conditions of the photo-resist on the substrate (gray layer) are colored differently. Blue, yellow and red areas stand for non-exposed, once exposed, and twice exposed photo-resist, respectively. If the irradiated energy of laser for the first exposure is higher than the threshold of development, the yellow and red areas dissolve in the development process. If the energy is slightly lower than the threshold, only the red area dissolves afterwards. Thus, one can prepare the remaining structure to be a dot matrix or a fishnet depending on the laser energy during the exposure. We demonstrated two types of structures by combination of WMIL and multi-exposing process. We prepared a waveguide on a substrate as designed in the inset of Fig. 3 (a). For the fabrication of dot matrix we expose a sample with the laser power of 7.3 $\rm mW/cm^2$ for 1 s. The actual resonance angle was $\theta$= 43.5$^\circ$ for this sample. The period of the structure estimated from the angle is 176 nm. The top view and birdfs-eye view of AFM images of the dot matrix structure developed by this condition are shown in Figs. 3 (c) and (e), respectively. The typical height of the dots in Fig. 3 (c) is 22 nm. 

For the fabrication of fishnet structures we decreased the laser power by 7 \%. The other conditions are completely the same as that of a dot-matrix structure. As a result, the complementary structure with a dot matrix was obtained as shown in Figs. 3 (d) and (f). The photo-resist layer is periodically perforated and looks like a fishnet. The typical depth of the hole in Fig. 3 (d) was 66.5 nm. These structures will be applicable to an etching mask with subwavelength structure in the visible region.

 \begin{figure}
\includegraphics[width=\linewidth]{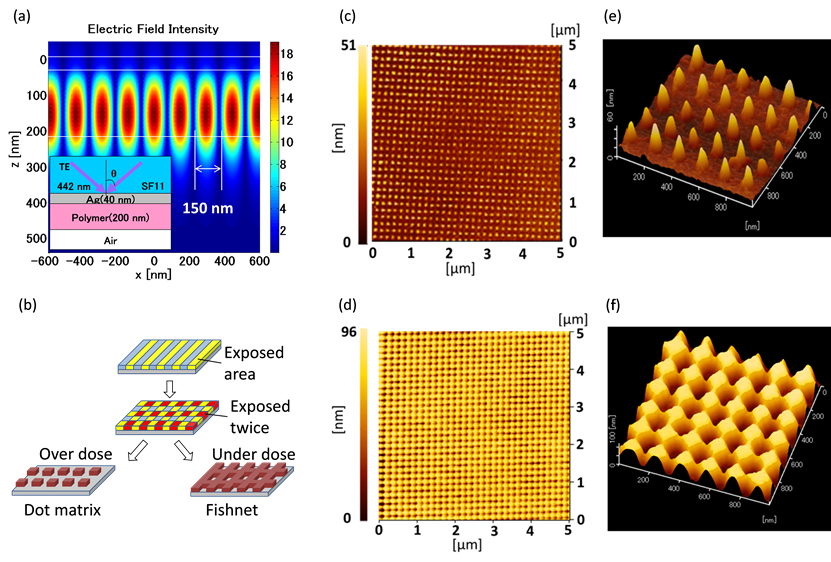}%
\caption{Two-dimensional structures on substrate. (a) Electric field distribution of TE1 mode in a waveguide shown in the inset. (b) Concept of process flow. Figures (c,e) and (d,f) are top and birdfs-eye views of the dot matrix structure and the fishnet structures, respectively.\label{fig3.png}}
 \end{figure}

\section{Three-dimensional structure}
WMIL can be utilized to fabricate three-dimensional structures by selecting an appropriate higher order waveguide mode, as have already been pointed out in Ref. [9]. Here we show that a complex field distribution along the z-direction can be realized even in  a TM1 mode. It is important to understand that photochemical reaction occurs in photoresist interact through the electric component of the light field. We demonstrate to fabricate a three-dimensional  structure using TM1 mode. The electric intensity distribution of TM1 mode in a waveguide is simulated as shown in Fig. 4 (a) for a configuration shown in the inset. The incident angle for the resonance condition was $\theta$ = 56.0$^\circ$. One can find a checkerboard like pattern of the electric field intensity in the waveguide. In case a positive type resist is used for a dielectric waveguide, the red areas, where the electric field is more intense than the blue, would dissolve out in the developing process if the developer could readily go into these parts. 

In order to prove the idea, we surveyed the sample surface and we found an area where the first layer is partially peeled off as is shown in Fig.4(b). Figure 4 (c) shows a magnified image of the part surrounded by red lines in Fig.4(b) where one can see that the lines in the first layer is aligned between the lines in the second layer.  In order to achieve a three dimensional structure for such a field distribution, we have to rely on a partial incompleteness of the development, otherwise the structure would collapse in this case. But by constructing an appropriate field distribution with interconnecting voids, it would be possible to fabricate three dimensional structure based on the WMIL technique.

 \begin{figure}
\includegraphics[width=\linewidth]{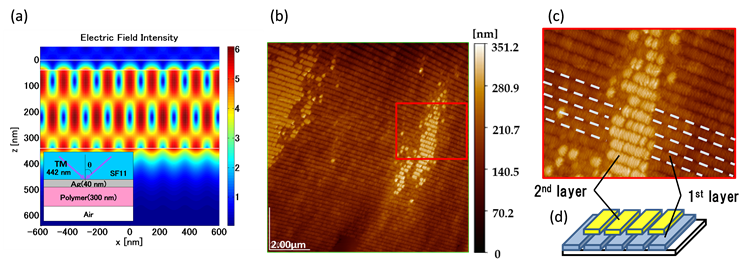}%
\caption{Three-dimensional structure. (a) Simulated field distribution for TM1 on the condition shown in the inset. (b) AFM image of a fabricated structure. (c) Magnified image inside a red rectangle in (b). Dashed lines are eye-guide drawn on the grooves of first layer. (d) The schematic image of this layered grating structure.\label{fig4.png}}
 \end{figure}

\section{Multi-layered coupler}
So far we have discussed WMIL with a single silver layer as a high reflective coupler to the resist layer. By replacing it with a Ag/dielectric/Ag trilayers, we propose to fabricate cut wire pair structures [12] as well as double fishnet structures [3] just by following the procedure in section 4. Here the Ag/dielectric/Ag layer serves as a coupling layer first and then is fabricated into metamaterials after development and etching processes. 

We designed a trilayered coupler as shown in Fig. 5 (a).  It is composed of Ag, Al2O3, and Ag layers all of which thickness was 30 nm.  The electric field distribution of the resonating TE1 mode with an incident angle of $\theta$ = 55.3$^\circ$ can be seen within the polymer part of Fig. 5 (b). We sputtered Ag/Al2O3/Ag layers before spin-coating a 200nm-thick photo-resist layer. After the WMIL process, we were able to fabricate a sample in fact as an AFM image is shown in Fig. 5 (c). A periodic structure is clearly seen on the trilayered surface.  If we prepare a two-dimensional resist pattern as a mask, it will be able to fabricate a double fish net structure by ething followed by resist removal. The trilayer can be extended even further to metal dielectric mulilayer, which can be regarded as hyperbolic metamaterials [13] and which will be applicable to multi-fishnet structures [14].

 \begin{figure}
\includegraphics[width=\linewidth]{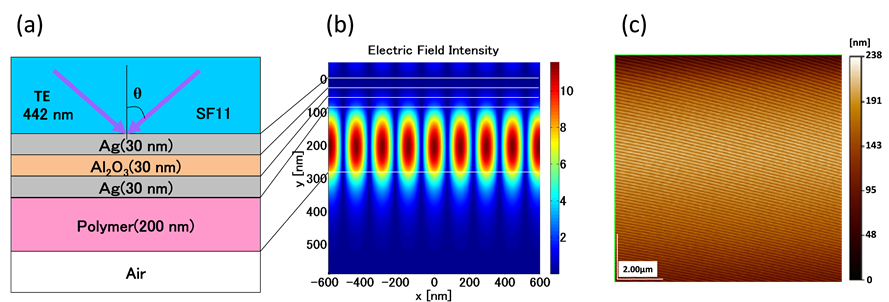}%
\caption{Example of a waveguide with trilayer coupler. (a) schematic structure. (b) Simulation results in the case at the resonance angle for TE1. (c) AFM image of fabricated resist structure on multilayer. 
.\label{fig5.png}}
 \end{figure}

\section{Application of WMIL to a metamaterial with LC resonance in the visible region}
In the end of this paper, we show that the split-ring resonator (SRR)-like structure can be realized by combining WMIL and oblique evaporation of metal. We start with a design shown in Fig. 6 (a). This structure is based on the multi-dielectric waveguide already mentioned in the previous section. By sputtering Ag obliquely on the resist structure from the upper right side, namely $\theta < 0$, of the figure, an array of isolated curved metallic objects is formed if the angle is large enough. We expect that this gone-side roof structure (OSRS)h on the resist grating has an LC resonance similar to SRRs. Because the cross-section of the OSRS in the y-plane is small and the gap is widely open compared to the conventional SRR structure [15], a LC resonance may be found in the visible range of the spectrum. Recently a design advantage of a split tube structure for magnetic resonance in the visible was pointed out [16]. Unlike the conventional inplane thin SRR, the split tube suffers from less electron scattering at the interface, which results in low loss. According to our numerical calculation with a commercial electromagnetic simulation software (CST Microwave studio), we found a resonance at around 2 eV. In Fig. 6 (b) we show $1-T-R$ obtained for TM-polarized light, where $T$ and $R$ are transmittance and reflectance, respectively. Considering the period of this structure of 165 nm, $1-T-R$ corresponds to absorption $A$ at the resonance as it is below the diffraction limit. Since the energy of this absorption band is independent of the incident angle, the absorption band is ascribed to a localized mode. Indeed, simulation shows that the electric field and the magnetic field are localized around the resonator, exchanging their energy at the resonant frequency. We also confirmed in the simulation that the coupling layer on the substrate have little influence on the band energy in this system. By passing, it is notable that the angle dependence of the absorption intensity is asymmetric due to the asymmetric structure. 

In order to fabricate such a metamaterial, we first prepared a grating structure with the WMIL using TE1 mode, as we did in Fig. 2. The fabrication condition so far was the same with that of the sample shown in Fig. 2. Subsequently, we spattered Ag layer from $\theta$ = -18.4$^\circ$ for 500 s with the sputtering rate of 0.12 nm / s. We made a metamaterial sample on the area of 2 mm x 2 mm, which is sufficiently large for optical measurement.

In order to characterize the structure, we measured transmittance and reflectance in the fabricated sample as a function of incident light. For reflective measurement, we used a $\theta$ ? 2$\theta$ stage, with which we cannot observe the reflectance for the normal incidence. The obtained angle resolved absorption spectra is shown in Fig. 6 (c). One can find an asymmetric absorption band around 1.9 eV, corresponding to the feature in the simulation. The reason for considerably broader energy bandwidth of this absorption band in our experiment can be ascribed to the difference in the actual structure and the model, optical parameters of materials used in the simulation, or the distribution of structural parameters in a sample.

 \begin{figure}
\includegraphics[width=\linewidth]{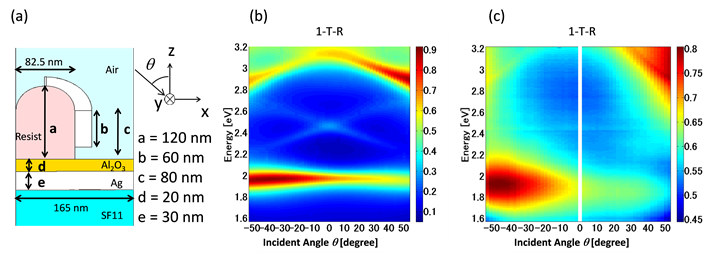}%
\caption{A metamaterial in the visible region. (a) Design of the structure based on the bilayer dielectric waveguide and the definition of coordinate for this system. Simulation (b) and experimental (c) results of angle resolved absorption spectra for TM polarized light in the metamaterial sample.
.\label{fig6.png}}
 \end{figure}

\section{Conclusion}
We investigated the various kinds of situation for WMIL toward the large area fabrication of subwavelength structures. By applying this technique to the waveguide whose thickness is comparable with the wavelength, high contrast and high aspect ratio can be achieved. Especially, the crisp distribution of TE1 mode along z-direction is useful to fabricate a well-defined subwavelength structure.

We succeeded to fabricate subwavelength structures with the multilayer waveguide design. Such a design allows us to choose a chemically unstable metal as a coupling layer. In case the refractive index of the dielectric is similar to that of a photo resist, the distribution of electric field is less affected from the insertion of the dielectric. On the other hand, this degree of freedom can be utilized to design the electric field in a waveguide. By introducing multi-layered coupling layer, we showed that large area resist ptterns for metamaterials, such as cut wire pair and double fishnet structure, can be prepared in a single or double exposure. Multiple exposure and the use of higher modes allow us to fabricate more complex structures. 

Oblique deposition of metal on the periodic resist patterns gives further variety, which may be utilized to fabricate SRR-like structure in the visible range of the spectrum. Thus we found that these features of WMIL are very suitable to fabricate metamaterials in the visible region.

\section*{ Acknowledgement}
The authors would like to thank Dr.\ Kei Sawada at SPring-8 Center, RIKEN, for fruitful discussion and comments from the theoretical point of view. The authors would also like to thank Mr.\ Makoto Saito at Tohoku University for his technical supports in experimental peripherals.

Hiroyuki Kurosawa is supported by the Japan Society for the Promotion of Science (JSPS), Grant-in-Aid for Young Scientists (Start-up) Grant Number 25889001 from KAKENHI. This work has been partially supported by the Grant-in-Aid for Scientific Research on Innovative Areas gElectromagnetic Metamaterials,h No. 22109005, from the Ministry of Education, Culture, Sports, Science and Technology (MEXT) of Japan.
\\

\end{document}